\documentclass[aps,twocolumn,english,prb,showpacs]{revtex4-1}

\usepackage{float}
\usepackage[dvips,final]{graphicx}
\usepackage{color}
\usepackage{amsmath}
\usepackage{graphics}
\usepackage{pslatex}
\usepackage{epsfig}
\usepackage{latexsym}
\usepackage{relsize}
\usepackage{bm}

\usepackage{xcolor}
\usepackage{soul}

\begin{document}
\title{Quantum Landau-Lifshitz-Bloch equation and its comparison with the classical case}

\author{P. Nieves$^{1}$, D. Serantes$^{1}$, U. Atxitia$^{2,3}$, and O.~Chubykalo-Fesenko$^{1}$ }
\affiliation{$^1$Instituto de Ciencia de Materiales de Madrid, CSIC, Cantoblanco, \mbox{28049 Madrid, Spain}}
\affiliation{$^2$ Fachbereich Physik, Universit\"{a}t Konstanz, D-78457 Konstanz, Germany}
\affiliation{$^3$ Zukunftskolleg, Universit\"{a}t Konstanz, D-78457 Konstanz, Germany}

\date{\today}

\begin{abstract}
The detailed derivation of the quantum Landau-Lifshitz-Bloch (qLLB) equation for simple spin-flip scattering mechanisms based on spin-phonon and spin-electron interactions is presented and the approximations are discussed.  The qLLB equation is written in the form, suitable for comparison with its classical counterpart. The temperature dependence of the macroscopic relaxation rates is discussed for both mechanisms. It is demonstrated that the magnetization dynamics is slower in the quantum case than in the classical one.
\end{abstract}
\pagebreak
\maketitle

\section{Introduction}
\label{sec:IntroductionLLBferrimagnet}

The Landau-Lifshitz-Bloch (LLB) equation has recently received a lot of attention as a high-temperature extension of the classical micromagnetism. \cite{Chubykalo,Lebecki} The use of the LLB-based micromagnetism is progressively becoming more  popular due to the  appearance of novel high-temperature magnetic applications.
The LLB formalism has been successfully used to model the heat-assisted magnetic recording, \cite{McDaniel,Finnochio} high-temperature spin-torque dynamics, \cite{Schieback} spin-caloritronics \cite{Hinzke} and  laser-induced magnetization dynamics. \cite{AtxitiaNi,Vahaplar} Apart from their fundamental interest,  these applications are very appealing from technological perspectives that range from energy saving strategies  to the increase of the speed of the magnetization switching.
Particularly, in the field of femtosecond optomagnetism, \cite{KirilyukRMP2010} where
a sub-ps demagnetization can be induced by the ultrafast heating produced by a femtosecond laser pulse,\cite{Ostler}
  the LLB equation has recommended itself as an useful approach. This is because it correctly describes the longitudinal magnetization relaxation in the strong internal exchange field, the key property of the magnetization dynamics at the timescale below 1 ps. \cite{AtxitiaNi,AtxitiaQ,Vahaplar}  Although the same characteristics have been proven to be reproduced by  atomistic  many-body  approach, \cite{Kazantseva}  the use of
the LLB micromagnetism for modeling purposes has some advantages: (i) the possibility to perform large scale modeling, for example, thermally-induced domain wall motion in much larger nanostructures  \cite{Hinzke} and (ii)  analytical  derivation of, for instance, the domain wall mobility \cite{quantumLLB} or the demagnetization time scales. \cite{AtxitiaQ, Nieves}

Up to now, most of works used the  classical version of the LLB equation
which was derived starting from a Heisenberg spin model and the Landau-Lifshitz equation for classical atomic spins. \cite{classicalLLB}  This has made the classical LLB approach very popular
since a direct comparison between
 the LLB and  the atomistic
simulations is therefore possible. \cite{Chubykalo, Kazantseva}
However,  the classical atomistic simulations mean effectively localized magnetic moments and correspond to the infinite spin number $S\rightarrow\infty$. As a consequence,  the magnetization versus temperature curve
  follows a Langevin function  rather than the Brillouin function which has been shown to fit
better for ferromagnetic metals,\cite{Culity} such as Ni and Co with $S = 1/2$, Fe with $S = 3/2$
 and Gd with $S = 7/2$.
In principle, the classical approximation seems hard to justify in the magnetic materials
 commonly used for ultrafast magnetization dynamics measurements, such as ferromagnetic metals, because
 of the delocalized nature of the relevant electrons responsible for the magnetic properties. However, recent works  which compare laser-induced magnetization dynamics  experiments  in metals with atomistic spin models\cite{Vahaplar,Radu,Ostler} as well as with their macroscopic counterpart - the classical LLB model\cite{Vahaplar, AtxitiaNi}- have  proven that both models are  very successful in the description and understanding of this phenomenon.

Similarly, the macroscopic three temperature model (M3TM), \cite{Koopmans,Schellekens}
has also been successfully  used in the description of \emph{femtomagnetism} experiments.
 The M3TM assumes a collection of  two level spin systems and uses a simple self-consistent Weiss mean-field model to evaluate the macroscopic magnetization. In the resulting system, importantly,
 the energy separation between levels is determined by a dynamical exchange interaction, similar to the LLB equation, which can be interpreted as a feedback effect to allow the correct account for the high temperature spin fluctuations. \cite{MuellerPRL2013}
This consideration turns out to be a fundamental ingredient for the correct description of the ultrafast
demagnetization in ferromagnets which suggests that the correct account for non-equilibrium thermodynamics  is probably more important than the correct band structure.
More recently, an alternative model to the M3TM and the LLB models, the so-called self-consistent Bloch (SCB) equation\cite{SCBloch} which
uses a quantum kinetic approach with the instantaneous local equilibrium approximation within
the molecular field approximation (MFA), has been suggested.

Both the M3TM and SCB models can account for the quantum nature of magnetism whereas the atomistic approach and the
classical LLB equation can not. However, the LLB model is not limited to the classical equation since there exists also
the quantum version of the LLB equation. The quantum LLB equation (qLLB)\cite{quantumLLB} has been derived even earlier than the classical one.\cite{classicalLLB} The derivation is based on the density matrix approach for the spin operators, similar to the SCB model,
and uses a dynamical exchange interaction within the MFA, similar to the  M3TM model.
One of the aims of the present paper is to study in more depth and to generalize the derivation of
the qLLB equation in order to clarify its use for the ultrafast dynamics. We also aim to show that it contains both
 SCB and M3TM equations for $S=1/2$.
Moreover, the qLLB equation has been barely investigated for numerical purposes. One of the reasons for that is mentioned above: the classical LLB equation allows the comparison with the atomistic simulations and, thus, its conclusions can be always checked. Another reason is the fact that the derivation has been made for the spin-phonon interaction mechanism which historically has been thought as the main  contribution to the magnetization damping. This mechanism is important for ps-ns applications at high temperatures such as spincaloritronics. Recent experiment also explore the possibility to excite  magnetization dynamics by acoustic pulses in picosecond range \cite{Scherbakov} (THz excitation) where the phonon mechanism is the predominant one.\cite{Kim}
  However, for the laser-induced magnetization dynamics where the spin-flips occur mainly due to the electron scattering, its relevance is marginal.  Thus, in this work we also derive the qLLB equation by considering a simple spin-electron interaction  as a source for magnetic relaxation.

  The article is organized as follows. In section II we briefly outline the qLLB derivation. The derivation of the qLLB equation above the Curie temperature as well as for the simplest electron-"impurity" mechanism are  presented. Comparatively to the original Garanin's derivation, \cite{quantumLLB} we discuss the approximations and put the qLLB equation in the form suitable for the comparison between the classical and the quantum cases. This allows us to relate the internal damping to microscopic scattering mechanisms. We  also show the equivalence of the qLLB equation for $S=1/2$ with the SCB and the M3TM models.   In section III we discuss the temperature dependence of the of the macroscopic longitudinal relaxation and the transverse damping as well as the internal microscopic coupling to the bath parameter within the two mechanisms. In section IV we present several numerical examples of the magnetization dynamics with the aim of comparison between the classical and the quantum cases. Finally, section V concludes the article and discusses possible extensions.

\section{Theoretical background for the quantum Landau-Lifshitz-Bloch equation}

\subsection{Basic assumptions for the qLLB equation with spin-phonon interaction}

For completeness and for subsequent development, in this first subsection of the paper we summarize the main aspects and approximations of the derivation of the qLLB equation. \cite{quantumLLB} The original derivation was done  assuming a magnetic ion interacting weakly with a thermal phonon bath via direct and the second order (Raman) spin-phonon processes.  The ferromagnetic interactions are taken into account in the mean-field approximation (MFA). The model Hamiltonian is  written as:
\begin{equation}
\hat{\mathcal{H}}=\hat{\mathcal{H}}_s+\hat{\mathcal{H}}_{ph}+\hat{V}_{\textit{s-ph}},
\label{eq:Hamiltonian}
\end{equation}
where $\hat{\mathcal{H}}_s$ describes the spin system energy, $\hat{\mathcal{H}}_{ph}$ describes the phonon energy, $\hat{V}_{\textit{s-ph}}$ describes the spin-phonon interaction:
\begin{eqnarray}
\label{eq:Vspin-phonon}
\hat{\mathcal{H}}_s & = & -\gamma  \mathbf{H}^{\textrm{MFA}} \cdot \hat{\textbf{S}}, \nonumber \\
\hat{\mathcal{H}}_{ph} & = & \sum_{q} \hbar \omega_{q} \hat{a}_{q}^{\dagger}\hat{a}_{q}, \\
\hat{V}_{\textit{s-ph}} & = & - \sum_{q}V_{q}(\mathbf{\boldsymbol{\eta}} \cdot \hat{\mathbf{S}})(\hat{a}_{q}^{\dagger}
+\hat{a}_{-q})-
\sum_{p,q}V_{p,q}(\mathbf{\boldsymbol{\eta}}\cdot\hat{\mathbf{S}})\hat{a}_{p}^{\dagger}\hat{a}_{q}.\nonumber
\end{eqnarray}
In the expressions above $\hat{\textbf{S}}$ is the spin operator, $\hat{a}^{\dagger}_{q}$ ($\hat{a}_{q}$) is the creation (annihilation) operator which creates (annihilates) a phonon with frequency $\omega_{q}$ where $q$ stands for the wave vector $\textbf{k}$ and the phonon polarization, and $\gamma = g\mu_B /\hbar$ is the gyromagnetic ratio where $g$ is the Land\'{e} g-factor, $\mu_{B}$ is the Bohr magneton and $\:\hbar$ is the reduced Planck constant.

The vector $\mathbf{H}^{\textrm{MFA}}$ is an effective field  in the MFA given by
\begin{equation}
\label{Hmfa}
 \mathbf{H}^{\textrm{MFA}}=\mathbf{H}_{E}+\mathbf{H}+\mathbf{H}_{K}=\frac{J_{0}}{\mu_{\textrm{at}}}\mathbf{m}+\mathbf{h},
 \end{equation}
where $\mathbf{H}_{E}=(J_{0}/\mu_{\textrm{at}})\textbf{m}$ is the homogeneous part of the exchange field, $J_0$ is the zero Fourier component of the exchange interaction related in the MFA to the Curie temperature $T_c$ as $J_0=3k_{B}T_cS/(S+1)$,   $\mu_{\textrm{at}}=g\mu_{B}S$ is the atomic magnetic moment, $\mathbf{m}=\langle \hat{\textbf{S}}(t)\rangle/\hbar S$ is the reduced magnetization where $\langle \ldots\rangle$ stands for the expectation value; and $\mathbf{h}=\mathbf{H}+\mathbf{H}_{K}$, where $\mathbf{H}$ is the external magnetic field and $\mathbf{H}_{K}$ represents the anisotropy field. Note that the original derivation\cite{quantumLLB} uses the two-site (exchange) anisotropy, since the treatment of the on-site anisotropy with a simple decoupling scheme, used below and suitable for the exchange interactions does not produce a correct  temperature dependence for the anisotropy. \cite{Bastardis} However, the on-site anisotropy can be later phenomenologically included into the consideration. \cite{classicalLLB,Kazantseva} Additionally, the inhomogeneous exchange field, $\propto (J_0 /\mu_{\textrm{at}} )\triangle \mathbf{m}$, may be either taken into account here or lately phenomenologically within the micromagnetic approach. \cite{Kazantseva}

The first term in the spin-phonon interaction potential $\hat{V}_{\textit{s-ph}}$ in Eq.\eqref{eq:Vspin-phonon} takes into account the direct spin-phonon scattering processes which are characterized  by the amplitude $V_q$, and the second term describes the Raman processes with amplitudes
$V_{p,q}$. The interaction may be anisotropic via the crystal field, which is taken into account through the parameter $\mathbf{\boldsymbol{\eta}}$.
 The spin-phonon scattering amplitudes $V_q$ and $V_{p,q}$ can be in principle  evaluated on the basis of the \emph{ab-initio} electronic structure theory.
Note that the interaction between spin and phonons considered in the Hamiltonian \eqref{eq:Hamiltonian} is one of the simplest possible forms, which has a linear (in the spin variable) coupling between spin and phonons. Based on the time reversal symmetry argument, it has been discussed\cite{Garanin4} that a quadratic spin-phonon coupling may be more physically justified. Nevertheless, it has been demonstrated that Eq. \eqref{eq:Hamiltonian} is adequate to describe the main qualitative properties of the spin dynamics.

The derivation of the qLLB equation \cite{quantumLLB} is based on a standard density matrix approach\cite{Blum,quantumLLB2} for a system interacting weakly with a bath. Namely, starting from the Schr\"{o}dinger equation one can obtain a Liouville equation for the time evolution of the density operator $\hat{\rho}=\vert\Psi\rangle\langle\Psi\vert$, where $\vert\Psi\rangle$ is the wave function of the whole system (spin and phonons). Next, the interactions with the bath are assumed to be small so that they can not cause a significant entanglement between both systems, this allows to factorize the density operator $\hat{\rho}$. Moreover, it is assumed that the bath is in thermal equilibrium (quasi-equilibrium) therefore, the density operator can be factorized by its spin and bath parts as $\hat{\rho}(t)\cong \hat{\rho}_{s}(t)\hat{\rho}_b^{eq}$, and after averaging over the bath variable one obtains the following equation of motion for the spin density operator $\hat{\rho}_s$\cite{quantumLLB2}
\begin{eqnarray}
\frac{\mathrm{d}}{\mathrm{d} t}\hat{\rho}_{s}(t)  = &  & \frac{i}{\hbar}\left[\hat{\mathcal{H}}_s,\hat{\rho}_{s}(t)\right]\nonumber\\
& - & \frac{1}{\hbar^{2}}\int_{0}^{t}dt'\textrm{Tr}_{b}\left[ \hat{V}_{\textit{s-ph}},\left[\hat{V}_{\textit{s-ph}}(t'-t)_{I}, \hat{\rho}_{s}(t'-t)_{I}\hat{\rho}^{eq}_{b}\right]\right], \nonumber\\
\label{eq:Liouville}
\end{eqnarray}
where $\textrm{Tr}_{b}$ is the trace over the bath variable, $\hat{V}_{\textit{s-ph}}(t'-t)_{I}=e^{-i(\hat{\mathcal{H}_s}+ \hat{\mathcal{H}}_{ph})(t'-t)/\hbar}\hat{V}_{\textit{s-ph}}e^{i(\hat{\mathcal{H}_s}+ \hat{\mathcal{H}}_{ph})(t'-t)/\hbar} $, $\hat{\rho}_{s}(t'-t)_{I}=e^{-i\hat{\mathcal{H}_s}(t'-t)/\hbar}\hat{\rho}_{s}e^{i\hat{\mathcal{H}_s}(t'-t)/\hbar} $, $\hat{\rho}_{s}(t)$ is written in terms of the Hubbard operators $\hat{X}^{mn}=\vert m\rangle\langle n\vert$ (where $\vert m\rangle$ and $\vert n\rangle$ are eigenvectors of $\hat{S}^{z}$, corresponding to the eigenstates $m\:\hbar$ and $n\:\hbar$, respectively), as
\begin{equation}
\hat{\rho}_{s}(t)=\sum_{m,n}\rho_{s,mn}(t)\hat{X}^{mn},
\label{eq:rho}
\end{equation}
where $\rho_{s,mn}(t)=\langle m\vert\hat{\rho}_{s}(t)\vert n \rangle$. Notice that in Eq. \eqref{eq:Liouville} time has been reversed ($t\rightarrow -t$) due to the definitions of $ \mathbf{m}$ and $ \hat{\mathcal{H}}_s $. Next, the following approximations are made: (i) the Markov or short memory approximation assuming that the interactions of the spins with the phonon bath are faster than the spin interactions themselves, this approximation means that in Eq. \eqref{eq:Liouville} the "coarse-grained" derivative is taken over time intervals $\Delta t$ which are longer than  the correlation time of the bath $\tau_b$ ($\Delta t\gg \tau_b$) and, (ii)  secular approximation, where only the resonant secular terms are retained, which consists in neglecting fast oscillating terms in Eq. \eqref{eq:Liouville}. It forces the time interval to be\cite{Blum} $\Delta t \gg\hbar/(E_{m}-E_{n})$ where $E_{m(n)}$ is an eigenvalue of $\hat{\mathcal{H}}_s$. For a ferromagnetic material with a strong exchange field $H_E$ we have $E_{m}-E_{n}\sim \hbar\gamma H_E$, therefore, for the Curie temperature  $T_c\simeq 800\;\textrm{K}$  we obtain $\Delta t \gg  1/\gamma H_E\sim 10\:\textrm{fs}$. Note that a different argument based on the scaling of the perturbation Hamiltonian (singular-coupling limit) can be found in Ref. \onlinecite{Breuer}. We should note that the validity of the above approximations for ultrafast magnetization processes may be questionable and should be checked in future on the basis of comparison with experiments. Note that similar studies for electronic coherence life time in molecular aggregates have found that the influence
 of the secular approximation in fs timescale is rather weak.\cite{secular} At the same time, the elimination of the secular approximation may be necessary for THz excitation of the spin system. On the other hand, if the Markov approximation is removed, it would meant an effective use of the colored noise. Our previous results \cite{AtxitiaPRL} indicate that the use of the colored noise with correlation time larger than 10 fs considerably slows  down the magnetization longitudinal relaxation time leading to time scales not consistent with those observed in experiments.

\begin{figure}[h!]
\centering
\includegraphics[width=\columnwidth ,angle=0]{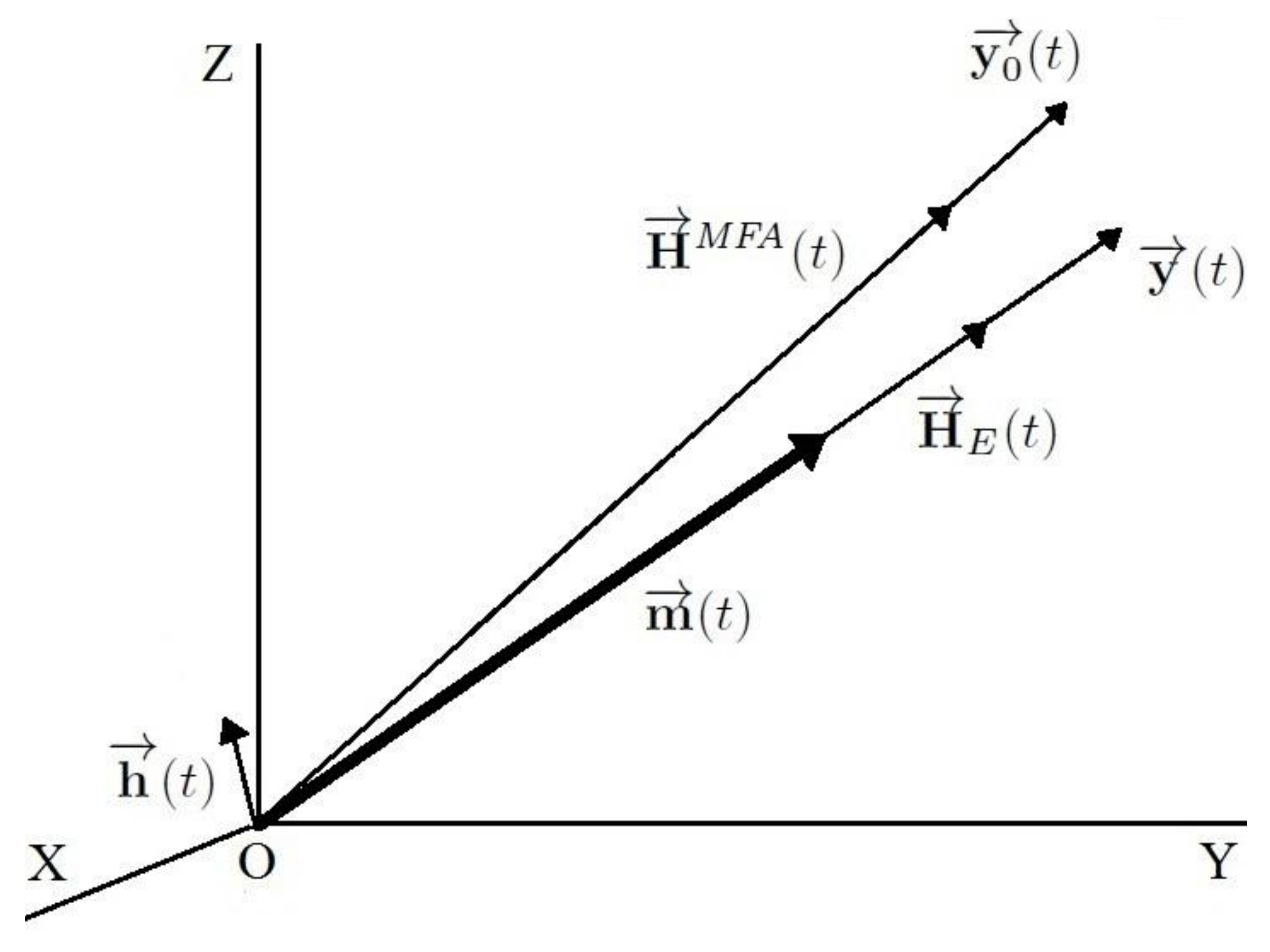}
\caption{Schematic diagram illustrating the relation between the reduced magnetization $\textbf{m}$ and the vectors $\textbf{H}_{E}$, $\textbf{y}$, $\textbf{h}$, $\textbf{H}^{\textit{MFA}}$ and $\textbf{y}_{0}$ at instant $t$ in a non-equilibrium state.}
\label{fig:diagram}
\end{figure}

As a result of these assumptions, one arrives to the equation for the Hubbard operators in the Heisenberg representation which for the isotropic case ($\eta_{x}=\eta_{y}=\eta_{z}=1$) becomes\cite{quantumLLB}
\begin{eqnarray}
& & \frac{\textrm{d}}{\textrm{d}t}\hat{X}^{mn}(t)  =  i\gamma H^{\textrm{MFA}}(m-n)\hat{X}^{mn}(t)-W_{1}(m-n)^{2}\hat{X}^{mn}(t)\nonumber\\
& - & W_{2}\Big\lbrace \frac{1}{2}\left[l^{2}_{m}+l_{n}^{2}+e^{-y_0}(l^{2}_{m-1}+l^{2}_{n-1})\right]\hat{X}^{mn}(t)\nonumber\\
&-& l_{m-1}l_{n-1}\hat{X}^{m-1,n-1}(t)-e^{-y_0}l_{m}l_{n}\hat{X}^{m+1,n+1}(t) \Big\rbrace ,
\label{eq:Xmn_ph}
\end{eqnarray}
where $\hat{X}^{mn}(t)=e^{-i\hat{\mathcal{H}_s}t/\hbar}\hat{X}^{mn}e^{i\hat{\mathcal{H}_s}t/\hbar}$, $y_0=\beta\: \hbar\: \gamma H^{\textrm{MFA}}$, $l_m =\sqrt{(S-m)(S+1+m)}$, $\beta = 1/k_B T$,
\begin{eqnarray}
W_{1}=\sum_{q,p}\vert V_{p,q} \vert^{2}n_{p}(n_{q}+1)\pi\delta(\omega_{q}-\omega_{p})
\label{w1}
\end{eqnarray}
\begin{eqnarray}
W_{2} & = & \sum_{q}\vert V_{q} \vert^{2} (n_{q}+1)\pi \delta (\omega_{q}-\gamma H^{\textrm{MFA}})\nonumber\\
& + & \sum_{p,q}\vert V_{p,q} \vert^{2} n_{p}(n_{q}+1)\pi \delta (\omega_{q}-\omega_{p}-\gamma H^{\textrm{MFA}}),
\label{w2}
\end{eqnarray}
 and $n_{q}=[\exp(\beta\hbar \omega_{q})-1]^{-1}$ is the Bose-Einstein distribution. Using Eq. \eqref{eq:Xmn_ph} and the relation between the spin operators $\hat{S}^{z}$, $\hat{S}^{\pm}\equiv \hat{S}^{x}\pm i\hat{S}^{y}$ and the Hubbard operators given by
\begin{eqnarray}
& & \hat{S}^{+}  =  \hbar\sum_{m=-S}^{S-1}l_{m}\hat{X}^{m+1,m}\:\: ,\:\:\hat{S}^{-}=\hbar\sum_{m=-S}^{S-1}l_{m}\hat{X}^{m,m+1}\: ,\nonumber\\
& & \hat{S}^{z}  =  \hbar\sum_{m=-S}^{S}m\hat{X}^{mm},
\label{S_hubbard}
\end{eqnarray}
one obtains a set of coupled equations of motion for the spin component operators which after averaging becomes
\begin{eqnarray}
\frac{d}{dt}\langle\hat{S}^{x(y)}\rangle & = & \mp \gamma H^{\textrm{MFA}}\langle\hat{S}^{y(x)}\rangle - (K_{1}+K_{2})\langle\hat{S}^{x(y)}\rangle\nonumber\\
& - & K_{2}\tanh\left(\frac{y_0}{2} \right)  \langle\hat{S}^{x(y)}\hat{S}^{z}+ \hat{S}^{z}\hat{S}^{x(y)}\rangle\label{DSx}\\
\frac{d}{dt}\langle\hat{S}^{z}\rangle  =  -2 &K_{2}& \langle\hat{S}^{z}\rangle + 2K_{2}\tanh\left( \frac{y_0}{2}\right)
  \langle(\hat{S}^{x})^{2}+(\hat{S}^{y})^{2}\rangle\label{DSz}
\end{eqnarray}
where
\begin{eqnarray}
K_1 & = & W_1, \label{k1}\\
K_{2} & = & \frac{1}{2}\left(1+e^{-y_0} \right)W_2 \label{k2}.
\end{eqnarray}
 The decoupling of the Eqs. \eqref{DSx} and \eqref{DSz} is produced only in three special cases:\cite{quantumLLB} (i) for $S=1/2$ where one gets the Bloch equation, also called self-consistent Bloch equation in Ref. \onlinecite{SCBloch} (also see below the subsection II.D) (ii) at high temperatures  ($k_{B}T \gg \hbar\gamma  H^{\textrm{MFA}}$) where a different form of the Bloch equation is obtained and (iii) the classical ($S\gg 1$) and low-temperature limits ($k_{B}T \ll \hbar\gamma  H^{\textrm{MFA}}$) where one obtains the Landau-Lifshitz-Gilbert equation (LLG). For the general case where the decoupling is not possible one can use the method of the modeling distribution functions, \cite{garanin3} assuming a suitable form for the spin density operator as follows
\begin{equation}
\label{distr}
\hat{\rho}_{s}(t)=\mathcal{Z}^{-1} \exp\left[\frac{\mathbf{y}(t) \cdot \hat{\mathbf{S}}}{\hbar}\right]\:\: ,\:\: \mathcal{Z}=\sum_{m=-S}^{S}\exp\left[y_0 m\right]
\end{equation}
where $\mathbf{y}(t)$ is an auxiliary dimensionless time-dependent function and its equilibrium value is $\mathbf{y}_{0}=\beta\gamma \ \hbar\textbf{H}^{\textrm{MFA}}$. It is possible to show
\cite{quantumLLB} that $\mathbf{y}(t)$ is related to the time-dependent reduced magnetization $\mathbf{m}(t)=\langle \hat{\textbf{S}}(t)\rangle/\hbar S$ as
\begin{equation}
\label{Eq_y}
\mathbf{m}(t)=B_S(Sy(t))\frac{\textbf{y}(t)}{y(t)},
\end{equation}
where $B_{S}(x)=[(2S+1)/2S]\coth ([2S+1]x/2S)-(1/2S)\coth (x/2S)$ is the Brillouin function for the spin value $S$. The spin operator averages in Eqs. \eqref{DSx} and \eqref{DSz} are calculated using the density matrix of the spin system given by Eq. \eqref{distr} as $\langle\hat{S}^{z}\rangle=\textrm{Tr}(\hat{\rho}_{s}\hat{S}^{z})$ and so on. Finally, after these calculations the Eqs. \eqref{DSx} and \eqref{DSz} have the following form in terms of the reduced magnetization\cite{quantumLLB}
\begin{widetext}
\begin{eqnarray}
\frac{d\mathbf{m}}{dt} & = & -\gamma\mathbf{m}\times\mathbf{h}
- K_{2}\frac{\tanh\left(\frac{y_{0}}{2}\right)}{\tanh\left(\frac{y}{2}\right)}\left(\frac{2(S+1)\tanh\left(\frac{y}{2}\right)}{m}-1\right)\frac{\mathbf{m}\times(\mathbf{m}\times\mathbf{h})}{mH^{\textrm{MFA}}}\nonumber \\
& - &
2K_{2}\left(1-\frac{\tanh\left(\frac{y_{0}}{2}\right)}{\tanh\left(\frac{y}{2}\right)}\frac{\mathbf{m}\cdot\mathbf{H}^{\textrm{MFA}}}{mH^{\textrm{MFA}}}\right)\mathbf{m} +  (K_{2}-K_{1})\left[\frac{(\mathbf{m}\times\mathbf{h})^{2}}{(mH^{\textrm{MFA}})^{2}}\mathbf{m+\frac{(\mathbf{m}\cdot\mathbf{H}^{\textrm{MFA}})\mathbf{m}\times(\mathbf{m}\times\mathbf{h})}{\textrm{\ensuremath{\mathrm{(\mathit{mH^{\textrm{MFA}}})^{2}}}}}}\right],
\label{eq:LLBexact}
\end{eqnarray}
\end{widetext}
where $y$ is defined through the relation Eq. \eqref{Eq_y}. In Fig.\ref{fig:diagram} we show the relation between the vectors $\textbf{m}$, $\textbf{H}_{E}$, $\textbf{y}$, $\textbf{h}$, $\textbf{H}^{\textrm{MFA}}$ and $\textbf{y}_{0}$ at instant $t$.

\subsection{Final form of the qLLB equation}

Eq. \eqref{eq:LLBexact} is not  convenient   for numerical  modeling or analytical considerations, since at each time step Eq. \eqref{Eq_y} should be solved to find the variable $\mathbf{y}(t)$ from $\mathbf{m}(t)$.
To avoid this issue,  we have to make further approximations, for instance, one can use that  in ferromagnets
the exchange field  is strong, $H_{\textrm{E}} \gg h$ in which case  $h/H_{E}$ is a small parameter.
Thus, in Eq.  \eqref{eq:LLBexact} only the terms linear in this parameter are retained. This assumption is valid both below $T_c$ (where always $H_{\textrm{E}} \gg h$) and close to $T_c$ where we can use the expansion
$H_{\textrm{E}} \simeq  (J_0/\mu_{\textrm{at}})( m_e + \widetilde{\chi}_{\parallel} h)$, where $m_e=B_S(\beta J_0 m_{e})$ is the equilibrium magnetization for $h=0$ and
%
\begin{equation}
\widetilde{\chi}_{\parallel}(T)=\left(\frac{\partial m}{\partial h}\right)_{h \rightarrow 0}
\end{equation}
is the reduced linear magnetic susceptibility. Since close to $T_c$ the susceptibility is large, $H_{\textrm{E}}\gg h$ for not too strong external magnetic fields.  Further simplification in Eq. \eqref{eq:LLBexact} is obtained using the fact that in stationary dynamic processes $y$ is close to the internal magnetic field direction, ($\vert y-y_0\vert \ll y$).\cite{quantumLLB} With these simplifications Eq. (\ref{eq:LLBexact}) is reduced to the qLLB equation in the form
\begin{eqnarray}
\frac{\mathrm{d} \textbf{m}}{\mathrm{d} t}
& = & -\gamma\textbf{m}\times\textbf{H}_\textrm{eff}
 +  \gamma\alpha_{\parallel} \frac{\textbf{m}\cdot\textbf{H}_\textrm{eff}}{m^{2}} \textbf{m}
 -\gamma\alpha_{\bot}\frac{\textbf{m}\times(\textbf{m}\times\textbf{H}_\textrm{eff})}{m^{2}},
 \nonumber \\
 \label{LLBfinalform}
\end{eqnarray}
where $\textbf{H}_\textrm{eff}$ is the effective field  given by
\begin{eqnarray}
\textbf{H}_\textrm{eff} & = &
\frac{1}{2\widetilde{\chi}_{\parallel}}
\left(1-\frac{ m^{2}}{m_{e}^{2}}\right)\textbf{m}+\textbf{h}\:\:  ,\: T<T_c.
 \label{heff}
\end{eqnarray}
The longitudinal susceptibility  $\widetilde{\chi}_{\parallel}$ can be evaluated in the MFA at $T< T_c$ as
$\widetilde{\chi}_{\parallel} =\mu_{at}\beta B'_S /(1-\beta  B'_S J_0)$ where $B'_S(x)=dB_S/dx$ is evaluated at the equilibrium
$B'_S=B'_S(\beta J_0 m_{e})$.
The parameters $\alpha_{\parallel}$ and $\alpha_{\perp}$  in Eq. \eqref{LLBfinalform} are the so-called longitudinal and transverse damping parameters, respectively.
In the present article we express them in a form  which is suitable for the comparison with the classical LLB equation. Below $T_c$ the damping parameters are written as
\begin{eqnarray}
\alpha_{\parallel} &=& \lambda \frac{2T}{3T_c}\frac{2q_{s}}{\sinh\left(2 q_{s} \right) }  \label{Longdamp}\\
\alpha_{\perp} & =&  \lambda\left[  \frac{\tanh(q_{s})}{q_{s}}-\frac{2T}{3T_c}\left(1-\frac{K_1}{2K_2}\right) \right] ,  \label{transdamp}
\end{eqnarray}
where $q_{s}=3T_{c}m_{e}/(2(S+1)T)$ and
\begin{eqnarray}
\lambda = K_{2} \frac{(S+1)}{S}\frac{\mu_{at}}{\gamma k_{B}T}.
\label{lambda}
\end{eqnarray}
 In Eq. (\ref{LLBfinalform}) all terms are linear in parameter $h/H_{E}$. Consequently, in Eqs. (\ref{Longdamp})-(\ref{lambda}) the field $\textbf{H}^{\textrm{MFA}}$ in $K_1$ and $K_2$ can be evaluated at the equilibrium.
Note that for $S \rightarrow \infty$ and $K_1= K_2$, Eqs. (\ref{Longdamp}) and (\ref{transdamp}) turns to the damping expressions in the classical LLB equation. \cite{classicalLLB} This allows us to conclude that
 $\lambda$ represents the intrinsic (Gilbert) damping (coupling to the bath) parameter used in the many-spin atomistic
approach. Eq. \eqref{lambda}  therefore  relates
the microscopic damping and the  scattering probabilities  through Eqs. \eqref{w1},\eqref{w2}, \eqref{k1},\eqref{k2}.
The temperature dependence of the intrinsic damping is discussed in section III.

Close to $T_c$,  the effective field used in Eq. \eqref{LLBfinalform} and given by Eq. \eqref{heff}
is  not very convenient for numerical calculations since  $m_e\rightarrow 0$ and $\widetilde{\chi}_{\parallel}\rightarrow\infty$. To solve this issue  we expand the Brillouin function up to the third order in small parameter $x=\beta J_0 m_e$:  $B_S(x)\simeq ax/3-bx^{3}/45$ and its derivative as $B'_S(x)\simeq a/3-bx^{2}/15$ where $a=(S+1)/S$ and $b=([2S+1]^{4}-1)/(2S)^{4}$.
Thus,
\begin{eqnarray}
m_{e}^{2}  \simeq  \frac{5A_s}{3}\epsilon\quad & , & \quad
\widetilde{\chi}_{\parallel}=\frac{\mu_{\textrm{at}}\beta B'_S}{1-\beta  B'_S J_0}\simeq \frac{\mu_{\textrm{at}}}{J_{0}}\frac{1}{2\epsilon},\label{eq:me_sus_approx}
\end{eqnarray}
where $A_{s}=2(S+1)^{2}/([S+1]^{2}+S^{2})$ and $\epsilon=(T_c-T)/T_c$ is small close to $T_c$. Eq. \eqref{heff} can be rewritten as
\begin{eqnarray}
\textbf{H}_\textrm{eff}
& = & \frac{J_{0}}{\mu_\textrm{at}}\left(\epsilon-\frac{3 m^{2}}{5A_s}\right)\textbf{m}+\textbf{h} \:\:  ,\:\: \vert\epsilon\vert\ll 1.
\label{heff2}
\end{eqnarray}
 Above $T_{c}$ we also re-write the effective field  in terms of the longitudinal susceptibility at $T>T_c$, i.e., $\tilde{\chi}_{\parallel}=\mu_{at}T_c/[J_0(T-T_c)]$. This equation is obtained from Eq. (\ref{eq:me_sus_approx}) and the well-known property\cite{Yeomans} of the susceptibility close to $T_c$, $2\tilde{\chi}_{\parallel,T<T_c}(\epsilon)=\tilde{\chi}_{\parallel,T>T_c}(-\epsilon)$. Thus, above $T_c$ the effective field is written as
\begin{eqnarray}
\textbf{H}_\textrm{eff}
& = & -\frac{1}{\tilde{\chi}_{\parallel}}\left(1+\frac{3 T_c m^{2}}{5A_s(T-T_c)}\right)\textbf{m}+\textbf{h} \:\:  ,\: \frac{T_c}{T-T_c}\gg 1\nonumber\\
\label{heff3}
\end{eqnarray}
Note that although $\widetilde{\chi}_{\parallel}$ is divergent at $T_c$ as corresponds to the second-order phase transition, the internal fields are the same for any $T_c-\varepsilon$ and $T_c+\varepsilon$ insuring that under the integration of the
Eq. \eqref{LLBfinalform}, $\mathbf{m}(t)$ rests continuous through the critical point, as it should be.

On the other hand, in the region just above $T_{c}$, $q_{s}=0$ and $K_1\cong K_2$ (see section III), so that the damping parameters become approximately the same and equal to
\begin{eqnarray}
\alpha_{\parallel}= \lambda \frac{2T}{3T_{c}},\;\; \alpha_{\perp} = \lambda \frac{2T}{3T_{c}}\left[1+\mathcal{O}(\epsilon)\right],\quad  \frac{T_c}{T-T_c}\gg 1
\label{damp_ph_Tc}
\end{eqnarray}
where the dependence on the spin value $S$ is included implicitly through  $\lambda$
[see Eq. \eqref{lambda}].
For $S\rightarrow\infty$ and high temperatures where $K_1= K_2$ the classical LLB equation above $T_c$ is again recovered.

\subsection{The qLLB equation for the electron-"impurity" scattering}

In this section we derive the qLLB equation for a very simple model for the spin-electron interaction Hamiltonian - the electron-"impurity" scattering model proposed by B. Koopmans \textit{et al.} in Ref. \onlinecite{DallaLonga} and F. Dalla Longa in Ref. \onlinecite{Dalla} for the laser induced magnetization dynamics. The model assumes an instantaneous
 thermalization of the optically excited electrons to the Fermi-Dirac distribution.
The Hamiltonian considered here consists of a spin system which weakly interacts with a spinless electron bath and it reads
\begin{eqnarray}
\hat{\mathcal{H}}=\hat{\mathcal{H}}_s+\hat{\mathcal{H}}_{e}+\hat{V}_{\textit{s-e}},
\label{eq:HamiltonianE}
\end{eqnarray}
where $\hat{\mathcal{H}}_s$ is the energy of the spin system, $\hat{\mathcal{H}}_e$ stands for the electron bath energy and $\hat{V}_{\textit{s-e}}$ describes the spin-electron interaction energy,
\begin{eqnarray}
\hat{\mathcal{H}}_s & = & -\gamma \mathbf{H}^{\textrm{MFA}}\cdot\hat{\textbf{S}},\\
\hat{\mathcal{H}}_e & = & \sum_{\textbf{k}}\epsilon_{\textbf{k}}\hat{c}_{\textbf{k}}^{\dagger}\hat{c}_{\textbf{k}},\\
\hat{V}_{\textit{s-e}} & = & -\sum_{\textbf{k},\textbf{k'}}V_{\textbf{k},\textbf{k'}}(\hat{S}^{+}+\hat{S}^{-})\hat{c}_{\textbf{k}}^{\dagger}\hat{c}_{\textbf{k'}}.
\label{eq:Hamiltonian_spin_elec}
\end{eqnarray}
 Here $\hat{c}^{\dagger}_{\textbf{k}}$ ($\hat{c}_{\textbf{k}}$) is the creation (annihilation) operator which creates (annihilates) an electron with momentum $\textbf{k}$, $\epsilon_{\textbf{k}} = \hbar^{2}k^{2}/(2m_{el})$, $m_{el}$ is the electron mass, $V_{\textbf{k},\textbf{k'}}$ describes the scattering amplitude. The vector $\mathbf{H}^{\textrm{MFA}}$ is given by Eq. \eqref{Hmfa}.
 Note that  we have chosen for the spin-electron  interaction
 the  minimal model that can capture the main features of the physics involved in the magnetization dynamics.
 In a slightly more sophisticated approach the electron-phonon scattering may be also included, leading to the two-temperature model. \cite{Koopmans}
 More rigorous approach, the  \emph{sp-d} model,  allows the description of the ultrafast magnetization dynamics in
 magnetic semiconductors\cite{Lukas} and ferromagnetic metals.\cite{Manchon, Gridnev}

Following the same procedure as in the section II.A, we obtain
\begin{eqnarray}
& & \frac{d}{dt}\hat{X}^{mn}(t)  =  i\gamma H^{\textrm{MFA}}(m-n)\hat{X}^{mn}(t)\nonumber\\
& - &  W_2\Big\lbrace \frac{1}{2}\left[l^{2}_{m}+l_{n}^{2}+e^{-y_0}(l^{2}_{m-1}+l^{2}_{n-1})\right]\hat{X}^{mn}(t)\nonumber\\
&-& l_{m-1}l_{n-1}\hat{X}^{m-1,n-1}(t)-e^{-y_0}l_{m}l_{n}\hat{X}^{m+1,n+1}(t) \Big\rbrace
\label{eq:Xmn_el}
\end{eqnarray}
where
\begin{eqnarray}
W_{2}
& = & 2\pi\sum_{\textbf{k},\textbf{k'}} \vert V_{\textbf{k},\textbf{k'}}\vert ^{2}\widetilde{n}_{\textbf{k}}(1-\widetilde{n}_{\textbf{k'}})  \delta\left(\gamma H^{\textrm{MFA}}-\frac{\epsilon_{\textbf{k}}-\epsilon_{\textbf{k'}}}{\hbar}\right),
\label{eq:w2_el}
\end{eqnarray}
$\widetilde{n}_{\textbf{k}}=[\exp(\beta(\epsilon_{\textbf{k}}-\mu))+1]^{-1}$ is the Fermi-Dirac distribution and $\mu$ is the chemical potential. Comparing Eq. \eqref{eq:Xmn_el} and Eq. \eqref{eq:Xmn_ph} we can see that
this mechanism leads to the same formal form for the qLLB equation but with $W_1=0$. We notice that since $W_1=0$ we have $K_1 = 0$, and the damping parameters below $T_c$ are given by
\begin{eqnarray}
\alpha_{\parallel}& = &\lambda \frac{2T}{3T_{c}}\frac{2q_{s}}{\sinh\left(2 q_{s} \right) } \label{Longdamp_elec1}\\
\alpha_{\perp} & = & \lambda\left[  \frac{\tanh(q_{s})}{q_{s}}-\frac{2T}{3T_{c}} \right] \label{transdamp_elec1}.
\end{eqnarray}
Differently to the isotropic spin-phonon scattering qLLB equation, considered above, for the electron-"impurity" scattering qLLB equation in the region just above $T_c$ the damping parameters are not approximately the same, \emph{i.e.},
\begin{equation}
\alpha_{\parallel}= \lambda \frac{2T}{3T_{c}},\;\; \alpha_{\perp} = \lambda \frac{T}{3T_{c}}\left[1+\mathcal{O}(\epsilon)\right],\quad  \frac{T_c}{T-T_c}\gg 1. \label{transdamp_elec2}
\end{equation}
Note that this is  a consequence of the fact that the model \eqref{eq:Hamiltonian_spin_elec} assumes an anisotropic scattering. In the qLLB model with anisotropic phonon's scattering, defined by $\eta_{z}=\eta_{y}=0$ and $\eta_{x}=2$ we obtain the same result (with a different value of $K_2$).

We should point out that the temperature in the qLLB equation for the electron-"impurity" scattering corresponds to the electron bath temperature while for the spin-phonon scattering corresponds to the phonon bath temperature. Therefore, these results validate the coupling of the qLLB equation to the electron bath temperature in the modeling of ultrafast laser induced magnetization dynamics.

\subsection{The special case with $S=1/2$.}

In the case of $S=1/2$ we can get more simple forms of the qLLB equation. Indeed, in this case $m(t)=B_{1/2}(y(t)/2)=\tanh(y(t)/2)$ and $m_0(t)=B_{1/2}(y_0(t)/2)=\tanh(y_0(t)/2)$. Moreover, Eq. (\ref{eq:LLBexact}) can be further simplified assuming a strong exchange field ($H_E\gg h$) which implies
\begin{eqnarray}
\frac{\mathbf{m} \cdot  \mathbf{H}^{MFA}}{m \cdot H^{MFA}} & = & 1+O\left(\left[\frac{h}{H_{E}}\right]^2\right),\\
(\mathbf{m}\times\mathbf{H}^{MFA})^{2} & = & O\left(\left[\frac{h}{H_{E}}\right]^2\right),
\end{eqnarray}
and using the vectorial relation  $\textbf{a}\times(\textbf{b}\times\textbf{c})=\textbf{b}(\textbf{a}\cdot\textbf{c})-\textbf{c}(\textbf{a}\cdot\textbf{b})$, Eq.
 \eqref{eq:LLBexact} becomes
\begin{equation}
\label{SCB1}
\frac{\textrm{d} \mathbf{m}}{\textrm{d}t}=-\gamma\mathbf{m}\times\mathbf{h}-
(K_1+K_2)\mathbf{m}+\left[2K_2+(K_1-K_2)\frac{m}{m_0}\right]\mathbf{m}_0,
\end{equation}
where $\mathbf{m}_0=\tanh(y_{0}/2)\mathbf{H}^{\textrm{MFA}}/H^{\textrm{MFA}}$ and $K_1,K_2$ can be evaluated at equilibrium. In two special cases: (a) when $K_1= K_2$ or (b) for longitudinal processes only, i.e. for collinear $\textbf{m}$, $\textbf{m}_{0}$ and $\mathbf{H}^{\textrm{MFA}}$ this equation can be further simplified. In both cases the Eq. \eqref{SCB1} becomes
\begin{equation}
\label{SCB}
\frac{\textrm{d} \mathbf{m}}{\textrm{d}t}=-\gamma\mathbf{m}\times\mathbf{h}-\frac{\mathbf{m}-\mathbf{m}_0}{\tau_s},
\end{equation}
where $\tau_s=1/(2K_2)$ and the precessional term is zero for the case (b). Eq. \eqref{SCB} in Ref. \onlinecite{SCBloch} was called the self-consistent Bloch (SCB) equation.

For the case (b) of a pure longitudinal dynamics the Eq. \eqref{SCB1} becomes
\begin{equation}
\label{3TMa}
\frac{\textrm{d} m}{\textrm{d} t}=-\frac{m}{\tau_s}\left[1-\frac{\tanh\left(\frac{y_{0}}{2}\right)}{\tanh\left(\frac{y}{2}\right)}\right].
\end{equation}
Assuming as before that in dynamical processes the  deviations between $y$ and $y_0$ are small i.e. $\vert y-y_0\vert\ll y$ we approximate
\begin{equation}
\label{3TMb}
1-\frac{\tanh\left(\frac{y_{0}}{2}\right)}{\tanh\left(\frac{y}{2}\right)}= \frac{\tanh\left(\frac{y}{2}\right)}{\tanh\left(\frac{y_0}{2}\right)}-1+\mathcal{O}([y-y_0]^{2}),
\end{equation}
and replacing Eq. \eqref{3TMb} in Eq. \eqref{3TMa} one gets
\begin{equation}
\label{3TMc}
\frac{\textrm{d} m}{\textrm{d}t}=\frac{m}{\tau_s}\left[1-m\coth\left(\frac{y_0}{2}\right)\right].
\end{equation}
We notice that for the case of strong exchange field ($\vert\textbf{H}_{E}\vert\gg \vert\textbf{h}\vert$) and $S=1/2$ we can write $y_{0}/2\simeq \beta\gamma \ \hbar H_E/2 = mT_c/T$. Eq. \eqref{3TMc} is the same as used in the M3TM model,\cite{Koopmans} in which case ${\tau_s}$ is related to concrete Elliott-Yafet scattering mechanism.

\section{Temperature dependence of the relaxation parameters}

The two main parameters which define the properties of the macroscopic magnetization dynamics can be obtained by linearisation of the LLB equation. Namely, they are the longitudinal relaxation time
\begin{equation}
\tau_{\parallel}=\frac{\widetilde{\chi}_{\parallel}}{\gamma\alpha_{\parallel}},
\label{tau_long}
\end{equation}
 and the transverse relaxation time $\tau_{\bot}$, i.e. the characteristic time taken by the
transverse component of magnetization to relax to the effective field
$\mathbf{h}$ including the external field and the anisotropy contributions
\begin{equation}
\tau_{\perp}=\frac{m_e}{\gamma\, h \alpha_{\perp}}.
\end{equation}
The corresponding transverse relaxation term of Eq. (\ref{LLBfinalform}) below $T_c$ may be put in the more common form of the macroscopic LLG equation. For this instead of the normalization of magnetisaion to the total spin polarisation, one should use its normalisation to the saturation magnetization value, i.e. $M_e(T)$. The resulting equation is the same LLB one\cite{AtxitiaQ} but with a different damping parameters, called here $\alpha_{\textrm{LLG}}$.   
This  allows to link the transverse magnetization
dynamics described by the LLB equation  with the macroscopic (Gilbert-like) temperature-dependent damping
\begin{equation}
\tau_{\perp}^{-1}\propto\alpha_{\textrm{LLG}}=\frac{\alpha_{\perp}}{m_e}.  \label{alpha_llg}
\end{equation}
Note that while both $\alpha_{\parallel}$ and $\alpha_{\perp}$ are continuous through $T_C$, the parameters $\tau_{\parallel}$ and $\alpha_{\textrm{LLG}}$ diverge at $T_c$, corresponding to the critical behavior at the phase transition.
Next we consider some limiting cases for these characteristic parameters, for relatively low temperatures and temperatures close to $T_c$.

\subsection {Longitudinal relaxation time}

The longitudinal relaxation time fundamentally depends on the longitudinal
susceptibility, $\widetilde{\chi}_{\parallel}$ and the longitudinal damping parameter, $\alpha_{\parallel}$.
For the longitudinal susceptibility, using the expansions of the Brillouin function in the corresponding temperature regimes, we obtain

\begin{equation}
\label{quantum_susc}
\widetilde{\chi}_{||}\cong \frac{\mu_{at}}{k_{B}T_{c}}
\begin{cases}
\frac{T_c}{TS^{2}}e^{-\frac{3T_{c}m_{e}}{T(S+1)}}&\qquad T\ll \min(T_{c},\frac{T_{c}}{S}),\\
\frac{T}{9T_{c}}\left(\frac{S+1}{S}\right)&\qquad\frac{T_{c}}{S}\ll T\ll T_{c},\\
\frac{(S+1)}{6S}\frac{T_{c}}{(T_{c}-T)}&\qquad\frac{T_c}{T_c-T}\gg 1,\\
\frac{(S+1)}{3S}\frac{T_{c}}{(T-T_{c})}&\qquad\frac{T_c}{T-T_c}\gg 1.
\end{cases}
\end{equation}
Note that the region $T\ll \min(T_{c},\frac{T_{c}}{S})$ does not allow the transition to the classical case ($S\rightarrow \infty$).  This transition takes place only in the region  $T_{c}/S\ll T\ll T_{c}$, the latter condition can be satisfied for $S\gg1$ only. This means that for a given spin $S\gg 1$ the quantum case becomes
approximately the classical one only at temperatures $T\gg T_{c}/S$ (or more exactly $T \gg 3T_c m_e/2S$), this result is obtained from the analysis of the conditions in which the Brillouin function becomes approximately the  Langevin one. Using Eqs. (\ref{quantum_susc}) and  the asymptotic behavior of $\alpha_{\parallel}$ in the limiting cases,  the longitudinal relaxation time in the limiting cases is given by:
\begin{equation}
\label{tau_par}
\tau_{||}\cong \frac{\mu_{at}}{2 \gamma\lambda k_{B}T_{c}} \frac{S+1}{S}
\begin{cases}
\frac{T_{c}}{ T S}&\: T\ll \min(T_{c},\frac{T_{c}}{S}),\\
\frac{1}{3} \left[1+\left(\frac{S}{S+1}\right)\frac{T}{T_c}\right]&\:\frac{T_{c}}{S}\ll T\ll T_{c},\\
\frac{T_c}{2 (T_{c}-T)}&\:\frac{T_c}{T_c-T}\gg 1,\\
\frac{T_c}{T-T_{c}}&\:\frac{T_c}{T-T_c}\gg 1.
\end{cases}
\end{equation}
Note that our results are in agreement with the well-known relation, proposed by Koopmans {\em et al.} \cite{Koopmans} that the ultrafast demagnetization time scales with the ratio $\mu_{at}/T_{c}$. As we pointed out elsewhere, \cite{AtxitiaQ} the complete expression involves also the internal coupling to the bath parameter $\lambda$, defined by the scattering rate.
The two last lines in Eq. (\ref{tau_par}) describe the effect of the critical slowing down near the critical temperature.
Furthermore, the relaxation time decreases with the increase of the quantum number $S$. Note also that the longitudinal relaxation time is twice larger above $T_c$ than below $T_c$.
\begin{figure}[h!]
\centering
\includegraphics[width=\columnwidth ,angle=0]{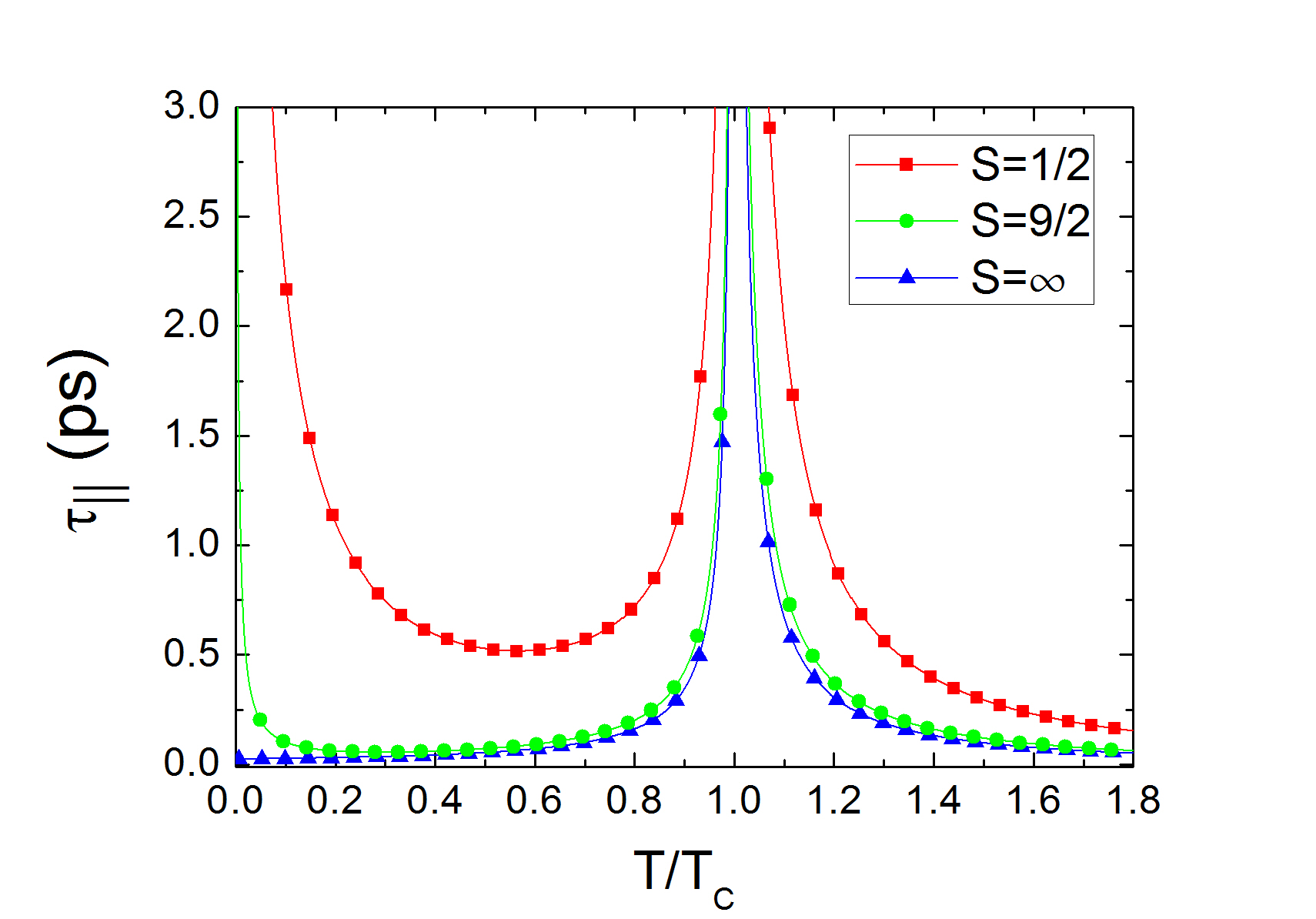}
\caption{Longitudinal relaxation time (Eq. \eqref{tau_long}) versus temperature using constant $\lambda=0.02$, $T_c=650 \:\textrm{K}$ and $\mu_{at}=0.5\mu_B$ in the three spin cases with $S=1/2$, $S=9/2$ and $S=\infty$. The case $S=\infty$ is done by taking the limit $S\rightarrow\infty$ in Eq. \eqref{tau_long}, which is equivalent to the classical  LLB equation.}
\label{fig:tau_long}
\end{figure}

Normally in the atomistic simulations one uses a constant in temperature coupling to the bath parameter $\lambda=$const. This gives the behaviour for the longitudinal relaxation time that we show in Fig. \ref{fig:tau_long} for the two limiting cases $S=1/2$ and $S=\infty$ and an intermediate case $S=9/2$. In the whole range of parameters the longitudinal relaxation slows down with the decrease of the spin value $S$.   For a finite spin number $S$ we observe a divergence of the relaxation time at low temperatures which does not happen for $S=\infty$. The intermediate case $S=9/2$ interpolates between a completely quantum case and a classical case. In this case, all asymptotic behaviors, described by  Eqs. (\ref{tau_par}) are observed, the longitudinal relaxation time diverges at low temperatures (as in the quantum case), is almost constant in the intermediate region (as in the classical case) and again diverges approaching to $T_c$.

The divergence of the longitudinal relaxation time at low temperatures  seems to be unphysical although it may be attributed to the freezing of the bath degrees of freedom and therefore, impossibility to absorb the energy from the spin system. One should note, however, that taking into account concrete physical mechanisms, the internal damping parameter $\lambda$ becomes temperature-dependent via Eq. (\ref{lambda}).

\begin{figure}[h!]
\centering
\includegraphics[width=\columnwidth ,angle=0]{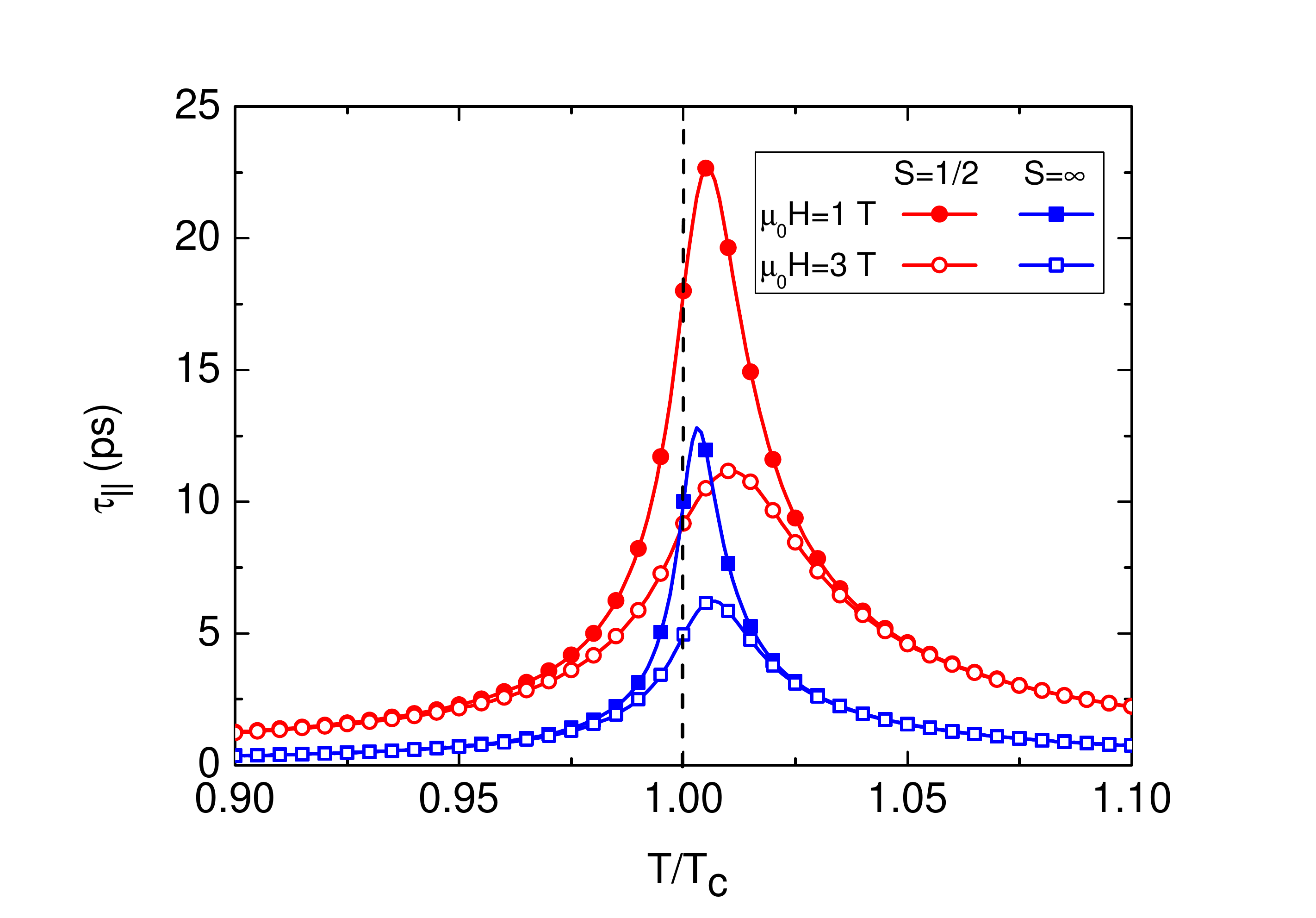}
\caption{The in-field longitudinal relaxation time calculated via direct integration of the qLLB equation with small deviation from the equilibrium. The following parameters are used $T_c=650$ K, $\mu_{\mathrm{at}}=0.5\mu_B$,  $\lambda =0.02$ and zero anisotropy constant. $\mu_0$ is the permeability of free space. }
\label{fig:susc_field}
\end{figure}

 In Fig. \ref{fig:susc_field} we present the longitudinal relaxation time as a function of the temperature in constant applied field for the two limiting cases $S=1/2$ and $S=\infty$. The longitudinal relaxation time was evaluated by direct integration of the qLLB equation with initial conditions $m_0-m_e=0.1m_e$.  The longitudinal relaxation time is smaller in the classical case than for the quantum one and, as expected, the maximum is displaced for larger values at larger fields. At $T\approx T_C$ the longitudinal relaxation time follows the expression
 \begin{equation}
 \tau_{||}(H,T=T_C)=\frac{5A_S \mu_{at}}{6\gamma \lambda J_0 m_H^2},\;\;m_H=\left(\frac{5A_S \mu_{at}H}{3J_0}\right)^{1/3}
 \end{equation}
 where $m_H$ is the field-induced equilibrium magnetisation at $T_c$.
  Therefore, unlike the statement of Ref. \onlinecite{SCBloch}, the in-field longitudinal relaxation time, calculated with LLB,  does not present any divergence at the Curie temperature.

\subsection{Transverse LLG-like damping parameter}

For the transverse damping we obtain the following limits
\begin{equation}
\label{alpha_LLG2}
\alpha_{LLG}\cong\lambda\begin{cases}
\frac{T}{3T_{c}}\left(2S+\frac{K_{1}}{K_{2}}\right)&\qquad T\ll \min(T_{c},\frac{T_{c}}{S}),\\
\left[1-\frac{1}{2S}-\frac{T}{3T_{C}}\left(1-\frac{K_{1}}{K_{2}}\right)\right]&\qquad\frac{T_{c}}{S}\ll T\ll T_{c},\\
\left(1+\frac{K_1}{K_2}\right)\sqrt{\frac{T_{c}}{15A_{S}(T_{c}-T)}}&\qquad\frac{T_c}{T_c-T}\gg 1.
\end{cases}
\end{equation}

\begin{figure}[h!]
\centering
\includegraphics[width=\columnwidth ,angle=0]{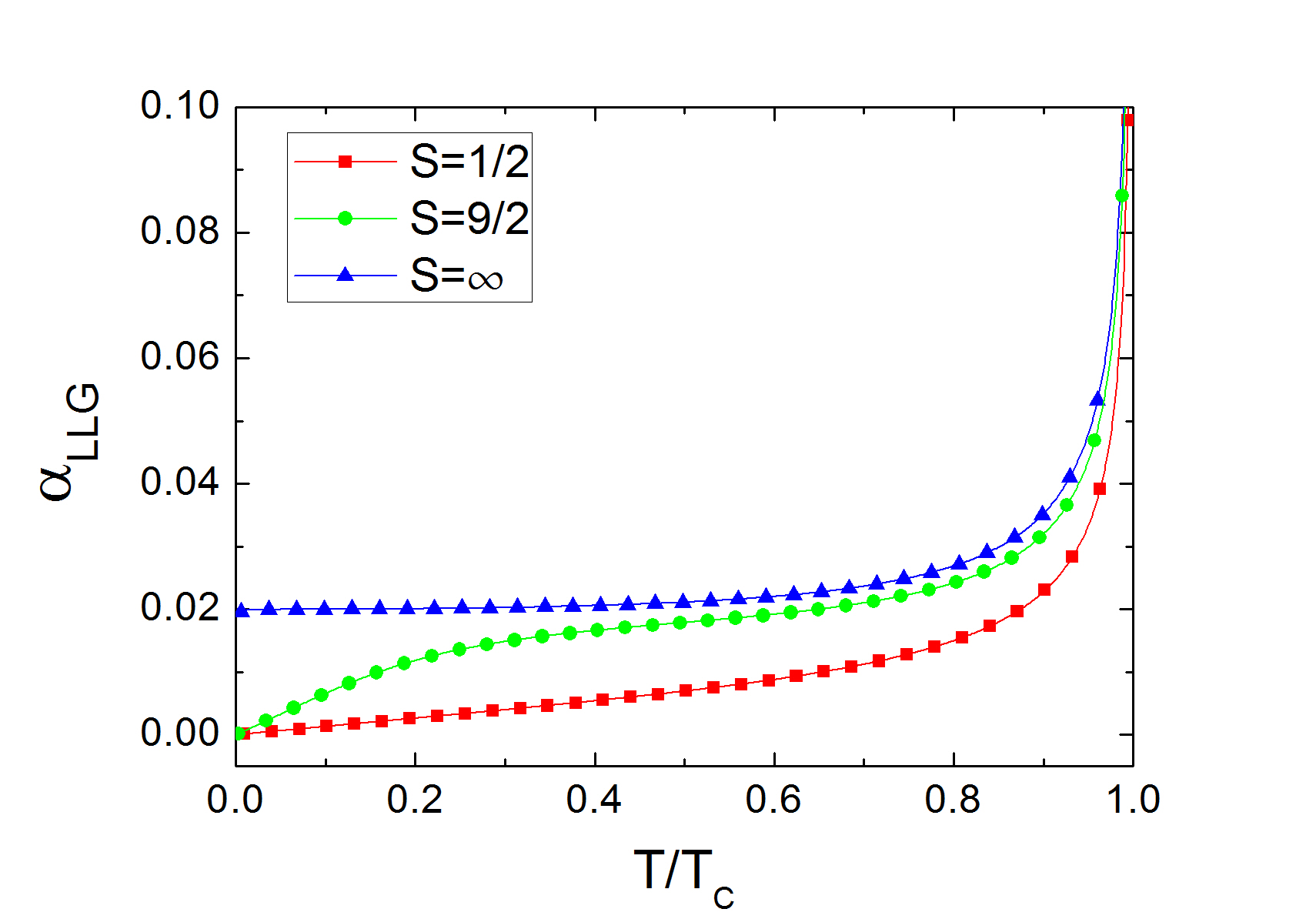}
\caption{LLG damping (Eq. \eqref{alpha_llg}) versus temperature using $K_1=K_2$, constant $\lambda=0.02$, $T_c=650 \:\textrm{K}$ and $\mu_{at}=0.5\mu_B$ for the three spin cases  $S=1/2$, $S=9/2$ and $S=\infty$.  The case $S=\infty$ is done by taking the limit $S\rightarrow\infty$ in Eq. \eqref{alpha_llg}, which is equivalent to the classical case.}
\label{fig:alpha_llg}
\end{figure}

 The temperature dependence of the LLG damping parameter for a constant value of $\lambda$ and $K_1=K_2$ is presented in Fig. \ref{fig:alpha_llg} for the two limiting cases $S=1/2$ and $S=\infty$ and the intermediate case $S=9/2$. In this case the transverse damping parameter tends to a constant value in the classical case and to a zero value in the quantum case. The transverse relaxation  also becomes faster with the increase of the spin number. For simplicity, we have used $\lambda=$const and $K_1=K_2$ in Fig. \ref{fig:alpha_llg} but, as we have seen before, the quantities $K_1$, $K_2$ and $\lambda$  depend on the particular scattering mechanism. Next we study the same limits but taking into account the  scattering mechanisms, considered here.

\subsection{Relaxation parameters with temperature-dependent internal scattering mechanisms}

\subsubsection{Scattering via phonons}

For the spin-phonon scattering we can evaluate $W_1$ and $W_2$ in Eqs. \eqref{w1}, \eqref{w2} using spin-phonon couplings of the type\cite{Garanin4}
\begin{equation}
V_q =\frac{\theta_{1}}{v}\sqrt{\frac{\omega_{q}}{M}}\quad ,\quad V_{pq}=\theta_2\frac{\sqrt{\omega_{p} \omega_{q}}}{Mv^{2}}
\end{equation}
where $\theta_1$ and $\theta_2$ are constants, $M$ is the unit cell mass and $v$ is the speed of sound in the material. The evaluation of $K_1$ and $K_2$ in Eqs. \eqref{k1}, \eqref{k2} gives the following result
\begin{eqnarray}
K_1\ll K_2 \simeq \frac{\theta_{1}^{2}\Omega(\gamma H^{\textrm{MFA}})^{3}}{4\pi Mv^{5}}\quad & , & \quad k_BT\ll\hbar\gamma H^{\textrm{MFA}}\label{k1_ph}\\
K_1\simeq K_2 \simeq \left[\frac{\theta_{2}^{2}\Omega^{2}k_{B}^{7}T_{D}^{5}}{20\pi^{3}M^{2}v^{10}\hbar^{7}}\right]T^{2}\quad & , & \quad  \hbar\gamma H^{\textrm{MFA}}\ll k_B T_D\ll k_BT\label{k2_ph}\nonumber\\
\end{eqnarray}
where $T_D$ is the Debye temperature and $\Omega$ is the unit-cell volume.  Using Eqs. \eqref{k1_ph}, \eqref{k2_ph} in Eq. \eqref{lambda} we obtain
\begin{equation}
\label{lambda_ph}
\lambda_{ph}\propto\begin{cases}
\frac{1}{T}&\qquad k_BT\ll\hbar\gamma H^{\textrm{MFA}},\\
T &\qquad \hbar\gamma H^{\textrm{MFA}}\ll k_B T_D\ll k_B T.
\end{cases}
\end{equation}
Therefore, if we take into account the temperature dependence of $K_1$, $K_2$ and $\lambda$ for the phonon scattering mechanism in Eqs. \eqref{tau_par} and \eqref{alpha_LLG2} we obtain
\begin{equation}
\label{tau_par_ph}
\tau_{||,ph}\propto
\begin{cases}
\textrm{const}&\quad T\ll \min(T_{c},\frac{T_{c}}{S})\: ,\: k_BT\ll\hbar\gamma H^{\textrm{MFA}}\\
T&\quad\frac{T_{c}}{S}\ll T\ll T_{c}\: ,\: k_BT\ll\hbar\gamma H^{\textrm{MFA}}\\
\frac{1}{T\vert T_{c}-T\vert}&\quad\frac{\vert T_c-T\vert}{T_c}\ll 1\: ,\: \hbar\gamma H^{\textrm{MFA}}\ll k_B T_D\ll k_B T,
\end{cases}
\end{equation}
and
\begin{equation}
\label{alpha_LLG2_ph}
\alpha_{LLG,ph}\propto\begin{cases}
\textrm{const}&\quad T\ll \min(T_{c},\frac{T_{c}}{S})\: ,\: k_BT\ll\hbar\gamma H^{\textrm{MFA}},\\
\frac{1}{T}&\quad\frac{T_{c}}{S}\ll T\ll T_{c}\: ,\: k_BT\ll\hbar\gamma H^{\textrm{MFA}}\\
\frac{T}{\sqrt{T_c-T}}&\quad\frac{T_c}{T_c-T}\gg 1\: ,\: \hbar\gamma H^{\textrm{MFA}}\ll k_B T_D\ll k_B T.
\end{cases}
\end{equation}

We observe that in the case of a pure phonon mechanism, the longitudinal relaxation time does not diverge at low temperatures. On the other hand, at elevated temperatures the longitudinal magnetization dynamics
is slowed down, since close to $T_c$ it is dominated by the divergence of $\widetilde{\chi}_{\|}$ [see Eq. \eqref{quantum_susc}] rather than by the  longitudinal damping parameter,  $\alpha_{\parallel}\propto T+\mathcal{O}({\epsilon})$. However, since $\tau^{-1}_{\bot}\propto \alpha_{\textrm{LLG}}$ we see that at high temperature the transverse magnetization dynamics becomes faster as the temperature gets closer to $T_c$.

\subsubsection{Scattering via electrons}

For the electron-"impurity" scattering we have found before that $K_1=0$, but we should still evaluate $K_2$. For this task, we calculate the quantity  $W_2$ given by Eq. \eqref{eq:w2_el} assuming that $\vert V_{\textbf{k},\textbf{k'}}\vert ^{2}=\mathcal{V}=
\textrm{const}$ and constant density of states around the Fermi level $D(\epsilon_F)$ (as in Refs. \onlinecite{DallaLonga} and \onlinecite{Dalla}). For this case, we obtain
\begin{eqnarray}
W_{2}
& = &\frac{\pi \mathcal{V}\hbar D(\epsilon_{F})^{2}}{2}\left[\frac{\hbar\gamma H^{\textrm{MFA}}}{1-e^{-\beta\hbar\gamma H^{\textrm{MFA}}}}\right],
\label{eq:w2_el2}
\end{eqnarray}
where $\epsilon_{F}$ is the Fermi energy,
\begin{equation}
D(E)=\frac{\Omega}{2\pi^{2}}\left(\frac{2m_{el}}{\hbar^{2}}\right)^{\frac{3}{2}}\sqrt{E}
\end{equation}
is the density of states for a free electron gas (taking into account the spin degeneracy) and $\Omega$ is the system volume. Replacing Eq. \eqref{eq:w2_el2} in Eq. \eqref{k2},  the following limiting cases for $K_2$ are obtained:
\begin{eqnarray}
K_2 & = & \frac{\pi \mathcal{V}\hbar^{2} D(\epsilon_{F})^{2}\gamma H^{\textrm{MFA}}}{4}\begin{cases}
1&\qquad k_BT\ll\hbar\gamma H^{\textrm{MFA}}\\
\frac{2k_B T}{\hbar\gamma H^{\textrm{MFA}}} &\qquad \hbar\gamma H^{\textrm{MFA}}\ll k_B T\nonumber\\
\end{cases},
\label{eq:k2b}
\end{eqnarray}
and then from Eq. \eqref{lambda} we obtain
\begin{equation}
\label{lambda_el}
\lambda_{el}\propto\begin{cases}
\frac{1}{T}&\qquad k_BT\ll\hbar\gamma H^{\textrm{MFA}},\\
\textrm{const} &\qquad \hbar\gamma H^{\textrm{MFA}}\ll k_B T.
\end{cases}
\end{equation}
We observe that at low temperatures ($k_BT\ll\hbar\gamma H^{\textrm{MFA}}$) $\lambda$ has the same temperature dependence as in the phonon scattering case, so that in this temperature regime we obtain the same results for $\tau_{||}$ and  $\alpha_{LLG}$ as in Eqs. \eqref{tau_par_ph} and \eqref{alpha_LLG2_ph}. However, at high temperatures ($\hbar\gamma H^{\textrm{MFA}}\ll k_B T$) we have $\lambda_{el} = \textrm{const}$, which validates the use of the constant $\lambda$ value in the modeling of the laser-induced magnetization dynamics, where the main mechanism is electronic and the temperatures are high. In this high temperature regime $\tau_{||}$ and  $\alpha_{LLG}$ have the following temperature dependencies
\begin{equation}
\label{tau_par_el}
\tau_{||,el}\propto
\frac{1}{\vert T_{c}-T\vert} \qquad\frac{\vert T_c-T\vert}{T_c}\ll 1\: ,\: \hbar\gamma H^{\textrm{MFA}}\ll k_B T,
\end{equation}
and
\begin{equation}
\label{alpha_LLG2_el}
\alpha_{LLG,el}\propto
\frac{1}{\sqrt{T_c-T}}\qquad\frac{T_c}{T_c-T}\gg 1\: ,\: \hbar\gamma H^{\textrm{MFA}}\ll k_B T.
\end{equation}
In this case, we also obtain a critical behavior of $\tau_{||}$ and  $\alpha_{LLG}$ close to $T_c$.

\section{Numerical comparison between classical and quantum cases}

In this section we compare the qLLB equation for $S=1/2$ and its classical limit ($S\gg 1$). We use the qLLB equation given by Eq. \eqref{LLBfinalform} for the isotropic phonon scattering mechanism and  the high temperature case ($K_1=K_2$). We note that for a proper comparison between classical and quantum cases, one should take the same magnetic moment $\mu_{\textrm{at}}$ and  Curie temperature $T_{c}$ (normally obtained from the experimental measurements) and not vary them with the spin number $S$. In the opposite case the magnetic moment would increase with $S$ and the Curie temperature decrease and the classical modeling results will not be recovered.

\begin{figure}[h!]
\centering
\includegraphics[width=\columnwidth ,angle=0]{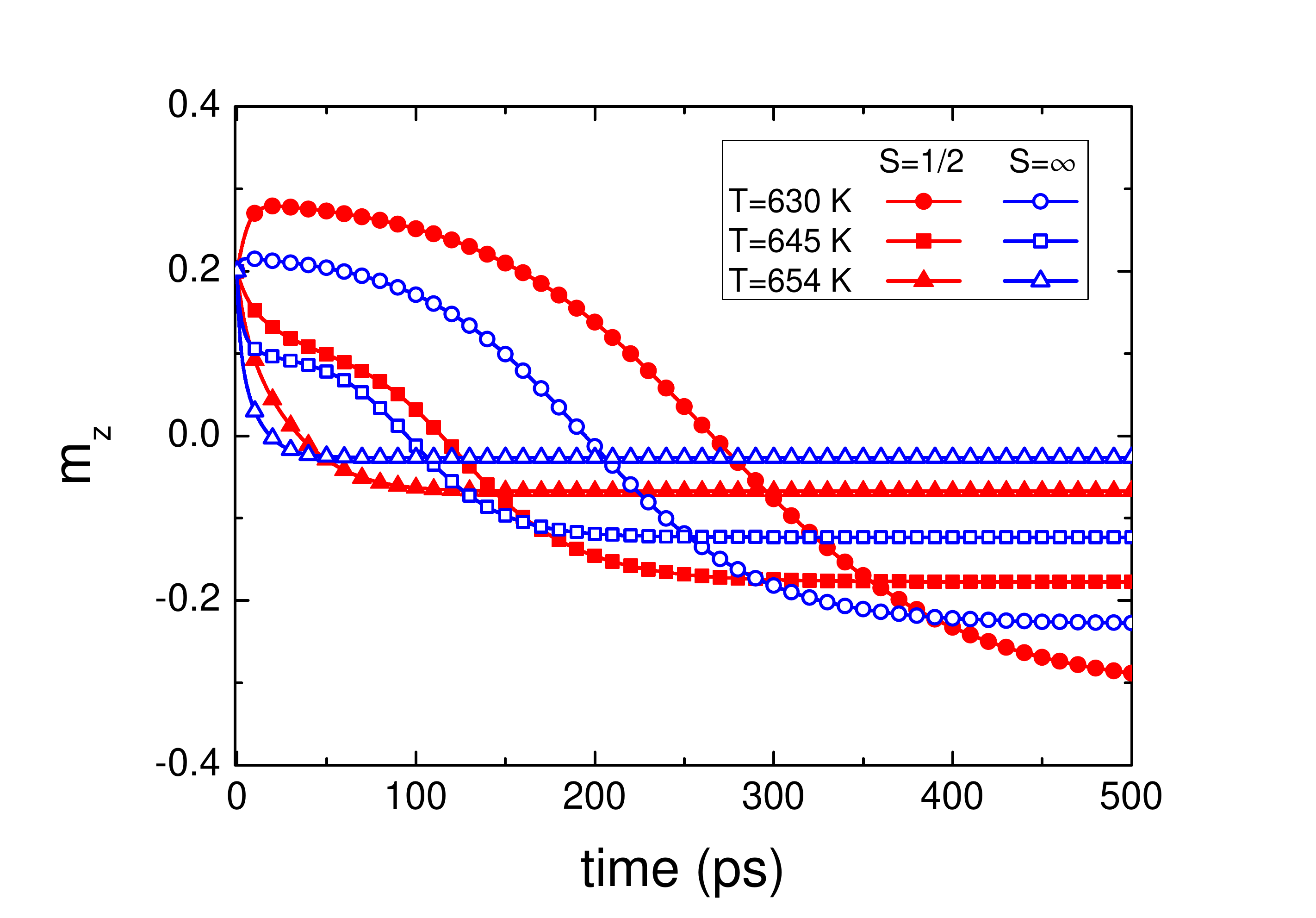}
\caption{The dynamics of $m_z$ component for the longitudinal plus transverse dynamics at $630$ K, $645$ K and $654$ K for $S=1/2,\infty$ where the initial magnetization is $\textbf{m}=(0.05,0,0.2)$ and the applied field is $\mu_0H_z=-1\textrm{T}$.}
\label{fig:mz_T}
\end{figure}

In our simulations we set $\gamma=1.76\times 10^{11}$ rad s$^{-1}$ T$^{-1}$, $T_c=650$ K, $\mu_{\mathrm{at}}=0.5\mu_B$ and $\lambda =0.02$ and zero anisotropy constant. Note that in order to be consistent with the  comparison of the SCB with $S=1/2$ (indistinguishable from the qLLB with $S=1/2$) and the classical LLB equation, presented in Ref. \onlinecite{SCBloch2},  we choose similar parameters and situations.
In Fig. \ref{fig:mz_T} we present the dynamics of $m_z$ component for $S=1/2,\infty$ and for different temperatures where the initial magnetization is set to $\textbf{m}=(0.05,0,0.2)$ and the external applied field is $\mu_0H_ z=-1\textrm{T}$, where $\mu_0$ is the permeability of free space. The initial response is slower for $S=1/2$ than for $S=\infty$ in agreement with the behavior of the longitudinal relaxation time, presented in Fig. \ref{fig:tau_long}. Note the variety of different functional responses and  that for the two cases below $T_c$ they cannot be represented as a one-exponential relaxation due to the nonlinearity of the LLB equation, prominent for $T$ close to $T_c$.

\begin{figure}[h!]
\centering
\includegraphics[width=\columnwidth ,angle=0]{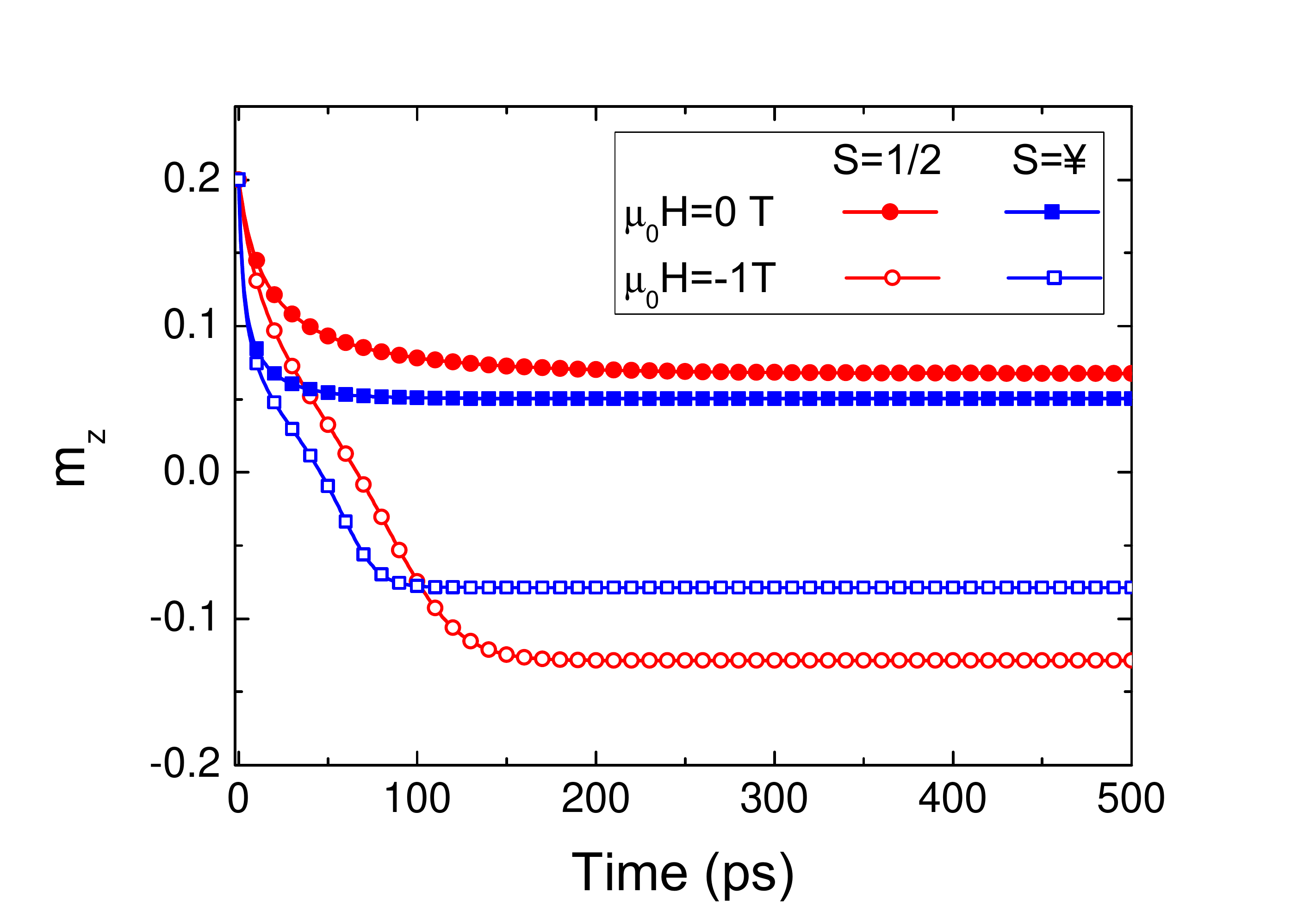}
\caption{The dynamics of $m_z$ component at $T=649$ K without and with external field $\mu_0H_z=-1\textrm{T}$  for $S=1/2,\infty$ where the initial magnetization is $\textbf{m}=(0.05,0,0.2)$.}
\label{fig:dyn_T649}
\end{figure}

In Fig. \ref{fig:dyn_T649} we present the relaxation of $m_z$ at $T=649$ K,  with and without an external field ($\mu_0H_z=-1\textrm{T}$) where the initial magnetization is set to $\textbf{m}=(0.05,0,0.2)$. We use the qLLB equation for $S=1/2$ and $S=\infty$ for comparison. Note that again the dynamics is faster for $S=\infty$ than for $S=1/2$. Since the qLLB and the SCB equations with $S=1/2$ are the same,  we conclude that the classical LLB equation gives a faster relaxation than the SCB equation, contrarily to the results  presented in Ref. \onlinecite{SCBloch2}.

\begin{figure}[h!]
\centering
\includegraphics[width=\columnwidth ,angle=0]{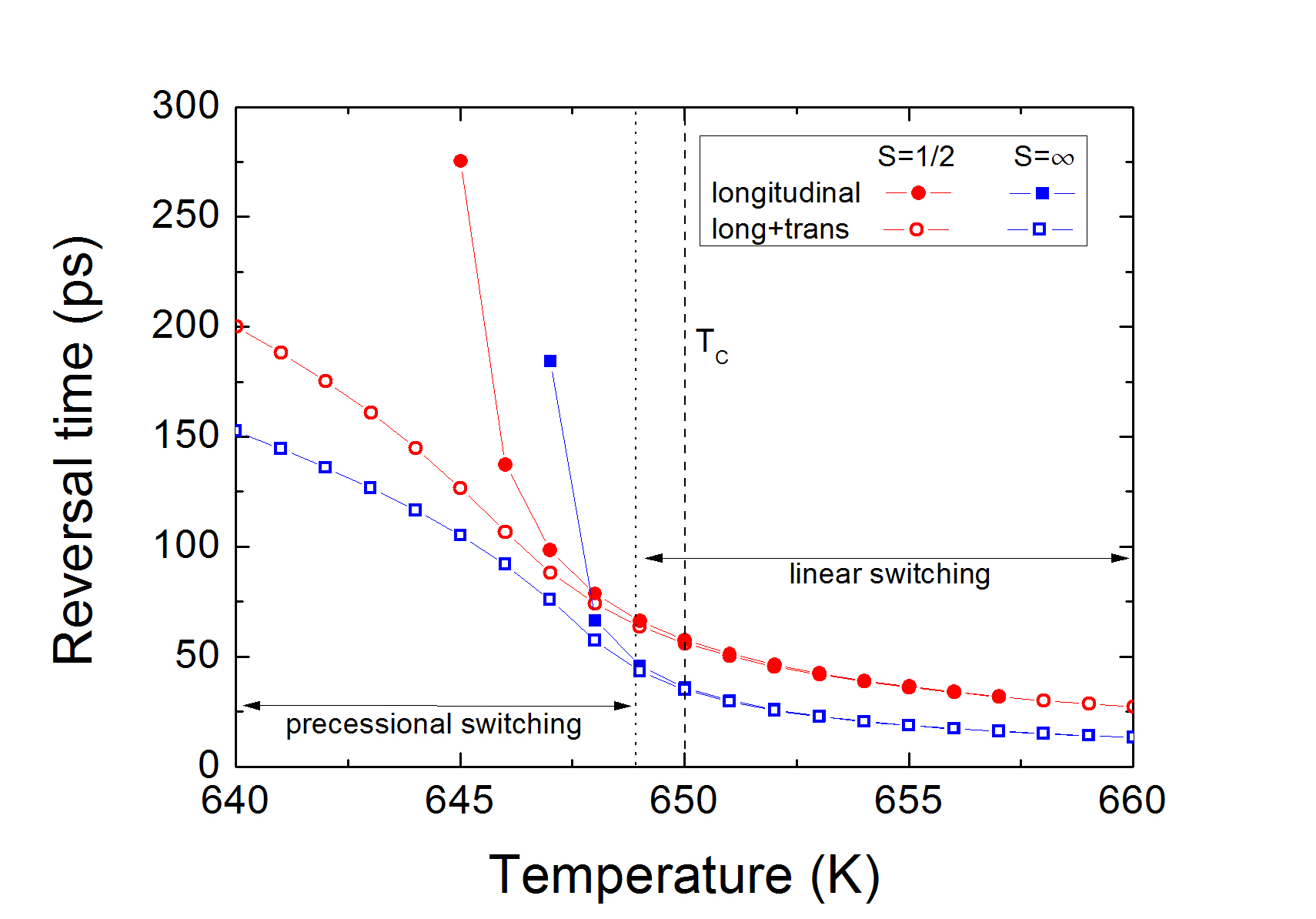}
\caption{Reversal time versus temperature for $S=1/2,\infty$ and $S=\infty$. In the pure longitudinal dynamics the initial magnetization is set to $\textbf{m}=(0,0,0.2)$ and in the longitudinal plus transverse dynamics the initial magnetization is set to $\textbf{m}=(0.05,0,0.2)$.}
\label{fig:rev_time}
\end{figure}

Similar to Ref. \onlinecite{SCBloch2} we define the reversal time as time elapsed between the initial  state and the instant of time at which the magnetization begins to reverse its direction, \emph{i.e.} crosses $m_z=0$ point. In Fig. \ref{fig:rev_time} we present the reversal time versus temperature for $S=1/2,\infty$ and for two different initial conditions: (i) pure longitudinal dynamics where the initial magnetization is set to $\textbf{m}=(0,0,0.2)$   and (ii) longitudinal plus transverse dynamics where the initial magnetization is set to $\textbf{m}=(0.05,0,0.2)$. We observe that the reversal time (for both the quantum and the classical case) does not present any discontinuity across the Curie temperature and is smaller for $S=\infty$ than for $S=1/2$, in contradiction to the results presented in Ref. \onlinecite{SCBloch2} where the SCB and the classical LLB equation were compared. As was pointed out in several previous publications, \cite{GaraninChubykalo,KazantsevaEPL,BarkerAPL} slightly below $T_c$ the magnetization reversal becomes linear, \emph{i.e.} occurs by a pure change of the magnetization magnitude. This path becomes not energetically favorable with the decrease of the temperature, the reversal path becomes elliptical and then completely precessional.

\section{Conclusions}
We have presented the derivation of the qLLB equation for two simple scattering mechanisms: based on the phonon and the electron-impurity spin-dependent scattering. While the spin-phonon interaction has been historically thought as the main contribution to the damping mechanism (for transverse magnetization dynamics), for
the ultrafast laser induced magnetization dynamics
the electron mechanism is considered to be the most important contribution.
At the same time, the induction of the ultrafast magnetization dynamics via acoustic excitation is becoming increasingly important so that the importance of the phonon-mediated mechanism is still relevant for femtomagnetism.
Although in the present work we have only considered the simplest form for the spin-phonon and -electron interaction Hamiltonian, the derivation could be generalized to more complex situations. The form of equation \eqref{LLBfinalform} is sufficiently general and at present can be used for modeling of most of the experimental cases,  understanding that the parameter $\lambda$ contains all necessary scattering mechanisms and can be extracted from experimental measurements as it was done before, \cite{AtxitiaNi,AtxitiaQ,Sultan,Koopmans} similar to the Gilbert damping parameter in standard micromagnetic modeling. Importantly,  the recently proposed self-consistent Bloch equation \cite{SCBloch,SCBloch2} and the M3TM model are contained in the qLLB model. \cite{Koopmans}

The derivation involves two important approximations: the Markov and the secular. Their validity could be questionable for the ultrafast processes and in the future these approximations should be investigated. At the same time, our comparisons with experiments for Ni,\cite{AtxitiaNi} Gd\cite{Sultan} and FePt\cite{FePt} have shown a very good agreement.

The derivation has allowed us to relate the classical internal coupling to the bath parameter $\lambda$, used in the atomistic spin model simulations,  to the scattering probabilities which could be evaluated on the basis of the \emph{ab-initio} electronic structure calculations, providing the route to a better scheme of the multi-scale modeling of magnetic materials.
The temperature dependence of $\lambda$ will depend on the nature of the concrete scattering mechanism.
In the present paper we have shown that  this parameter is temperature dependent.
At the same time,
the use of the temperature-independent microscopic damping (coupling to the bath parameter) for laser-induced magnetization dynamics, as it is normally done in the atomistic simulations, is probably reasonable. Our results also include the temperature dependence of macroscopic relaxation parameters: the longitudinal relaxation and the LLG-like transverse damping. We have shown that both transverse and longitudinal relaxation are faster in the classical case than in the quantum one.

The comparison between the classical and the quantum LLB equations has been done in the conditions of the same magnetic moment and the Curie temperature, as corresponds to the spirit of the classical atomistic modeling.  Unlike the statement appearing in Ref. \onlinecite{SCBloch2}, the magnetization is continuous when going through $T_{c}$, the same happens with the reversal time. In the considered case in this work, the reversal time is smaller in the classical case than in the quantum one, although our investigation shows that this result depends on the system parameters.

Our results contribute to a construction of correct multi-scale/micromagnetic approach for the modeling of high-temperature and/or short timescale magnetization dynamics. The obtained micromagnetic approach can be used for modeling of large structures, such as dots and stripes up to micron-sizes, under the conditions where the use of the LLB equation is necessary.

\section*{Acknowledgement}
This work was supported by  the Spanish Ministry of Economy and Competitiveness under the grant
FIS2010-20979-C02-02
and by the European Community's
Seventh Framework Programme (FP7/2007-2013) under
grant agreement No. 281043, FEMTOSPIN.
  U. A. acknowledges support from the EU FP7 Marie Curie Zukunftskolleg Incoming Fellowship Programme, University of Konstanz.


\end{document}